# Designing Interaction for Multi-agent Cooperative System in an Office Environment

Chao Wang, Stephan Hasler, Manuel Mühlig, Frank Joublin, Antonello Ceravola, Joerg Deigmoeller, Lydia Fischer

Honda Research Institute Europe GmbH
Carl-Legien-Straße 30, 63073 Offenbach am Main
{chao.wang,stephan.hasler,manuel.muehlig,frank.Joublin, joerg.deigmoeller,lydia.Fischer}@honda-ri.de

**ABSTRACT**

Future intelligent system will involve very various types of artificial agents, such as mobile robots, smart home infrastructure or personal devices, which share data and collaborate with each other to execute certain tasks. Designing an efficient human-machine interface, which can support users to express needs to the system, supervise the collaboration progress of different entities and evaluate the result, will be challengeable. This paper presents the design and implementation of the human-machine interface of Intelligent Cyber-Physical system (ICPS), which is a multi-entity coordination system of robots and other smart devices in a working environment. ICPS gathers sensory data from entities and then receives users' command, then optimizes plans to utilize the capability of different entities to serve people. Using multi-model interaction methods, e.g. graphical interfaces, speech interaction, gestures and facial expressions, ICPS is able to receive inputs from users through different entities, keep users aware of the progress and accomplish the task efficiently.

**CCS CONCEPTS**

• **Human-centered computing~Human computer interaction (HCI)**

## 1 Introduction

Multi-robots concept was introduced in the early 2000s to improve the system's robustness and capabilities [2]. After 20 years of development, the current multi-robot system becomes more complex and consists of multiple artificial agents. Those agents can be very different in their form and functionality, such as mobile robots, static smart home infrastructure, or smartphones. One of the challenges for an intelligent system can be seamless interactions between artificial agents and human, which requires the system share concepts about existing objects and ongoing events in their environment[11][12]. Our work presents a human-machine interface with which is aimed at tackling the mentioned challenge. The interface design is based on a multi-robotic system called ICPS (Intelligent Cyber-Physical system), which is implemented in a typical office workspace. It consists of three kinds of entities: SmartLobby, which is a lobby equipped with cameras and other sensors, and touch screen tables; Johnny, Ira and Walker, three mobile robots that are able to move inside the office; and Receptionist, a stationary booth at the reception of the office, which is equipped with camera, microphone and a touch screen. These entities are coordinated by ICPS to perform certain tasks, such as fetching objects, searching for persons or guiding guests to specific locations. Through Multi-model interaction methods, e.g. graphical interfaces, speech interaction, gestures and facial expressions, ICPS is able to receive requests from users, provide feedback about the progress and execute the task efficiently.

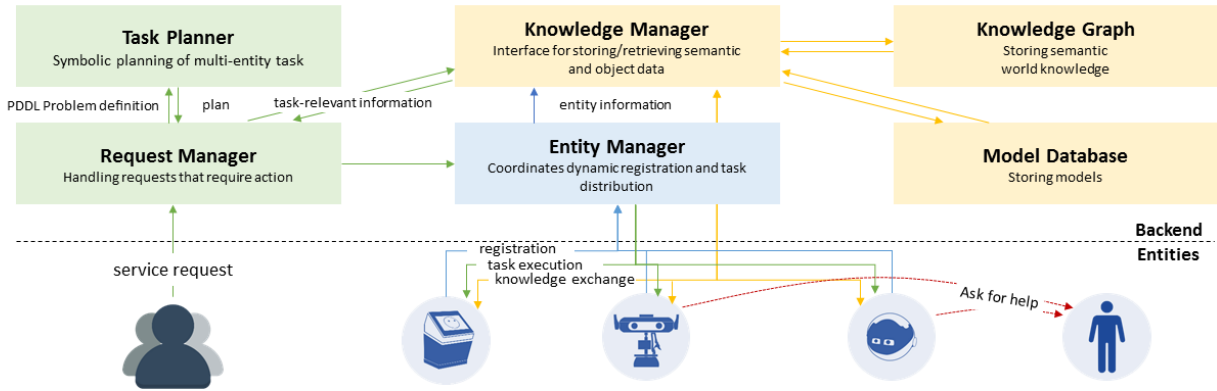

**Figure 1:** The system overview shows the backend parts and the entities. The green, yellow, and blue parts relate to planning, to storing and accessing the knowledge of the system, and to the registration of entities and the task distribution of created plans.

## 2 Related Work

There are multiple related publications in the domain of human-robot interaction in an office environment. For example, the CMU Snackbot project [8] implemented the industrial design of an autonomous robot for snack delivery in an office space targeted at long-term operation. Also, some researchers [13] showed a service robot system that focused on the longer operation and on asking humans for help. In STRANDS project [6], researchers implemented a robot to monitor an indoor office environment and generate alerts when they observed prohibited or unusual events. The robot has a head, eyes and led-lights, which can deliver non-verbal communication cues. Leonardi et al [9] suggested an interface which enables a native user to trigger certain actions based on personalized rules. The triggers include various IoT devices, such as wearables, lights or smart TV. The single robot architecture has been proven to be stable over a long time, pursuing service tasks in interaction with people. However, there is little research regarding multi-robot collaboration to service people in the long term running.

Since we are targeting at operating in a real office environment, and the present humans need to be considered by the system. A recent overview of the field of human-robot teaming and the associated challenges can be found in [3]. Also, there is earlier work considering humans during robots' actions [1], where the state of the human (e. g., standing or sitting) is considered for appropriate motion planning. We are however trying to more tightly incorporate the humans in the system's behaviour. As will be explained later in this work, humans are not seen as uncontrollable constraints, but instead, their capabilities are taken into account and the system might opt to ask the human for help.

One part of a multi-entity system is the knowledge organization and distribution amongst the components. This requires the right abstraction level due to the different sensors and capabilities of the system entities. Semantic representations are an efficient method to achieve this and a way to make knowledge gathered by single robots available to other robots [16]. As presented in [14], this seems feasible in a larger scope by using a representation that includes knowledge about the environment as well as past actions of a robot. Semantic representations also provide different ways of allowing for extendibility of the knowledge under an open-world assumption. On the one hand, it allows for reasoning over unknowns, for example by incorporating the concept of hypotheses [7]. On the other hand, the semantic representation can be connected to external world knowledge. This has for example been shown in the KnowRob project [15], where information from sensory data is associated to predefined ontological information. Furthermore, in our earlier work, we also showed that such a representation is well suited for interacting with humans and for generating human-understandable explanations of a reasoning process [4]. In that previous work, the focus was more on how to represent knowledge (in particular relations between tools, actions, and objects) rather than on symbolic planning. For planning, we are building upon traditional AI methods similar to the work presented in [5]. There, the planning domain and problem (e. g., a search task in an unknown environment) are defined using a variant of the Planning Domain Definition Language (PDDL) and a plan is searched for a single robot using a combination of deterministic and decision-theoretic planners. The process is similar in our work, besides that in our case, the planning problem is generated automatically according to which entities (and persons) are actually available and which capabilities they have.

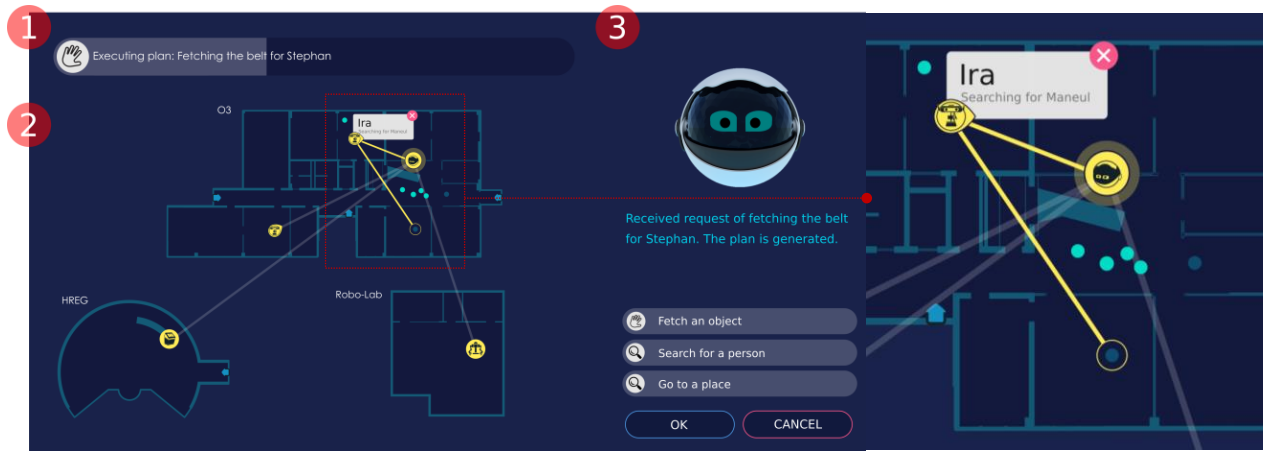

**Figure 2: The console interface of in the SmartLobby, which is displayed on the left touch table and big screen on the wall (see Figure 3). 1. Progress bar area: The icon in the progress bar shows the current goal; the text shows the current step that is executing; and the grey bar shows the percentage of the completion. 2. Maps area: this area shows maps of the three offices in Honda Research Institute, status and position of different entities. Right pictures shows details of the interface: The highlighted blue dots represent the active person; the dimmed blue dots represent the person detected in the last 5mins; the highlighted yellow lane shows the entities that are involved the current step of executing the task. For example, now robot Ira is going to a room to search for a person, and then ask him for the "belt" object. 3. Operation area. A virtual robot is shown in this area, which enables speech interaction. Besides, user can use shortcut buttons to send pre-set order to the system.**

## 3 System Design of ICPS

Figure 1 provides an overview of the components of the presented system. It is implemented as a centralized architecture including a backend and the entities that it controls. The term entity in this context is not restricted to robots, but also includes smart infrastructure, such as the depicted *SmartLobby*. Entities themselves do not communicate among each other, but only with two backend components. Firstly, the Entity Manager, which allows entities to register at the system and for the backend to assume control of them. Secondly, the Knowledge Manager, which coordinates the storage of sensory information received from the entities and allows other components to query this information via a common interface. For actually performing a function that utilizes the registered entities, the Request Manager receives the desired goal from a user. It then queries information relevant for the planning problem using the Knowledge Manager, such as which registered entities to use, their location and capabilities, and the task-relevant subset of the measured world state. This information is used to generate a sequential plan that is then executed via the Entity Manager. The system and the entities are implemented using ROS (Robert Operating System, see [10]) with multiple ROS masters to increase robustness wrt. Wireless communication outage (based partially on the multimaster_fkie package[1]). The subsequent section provides more details on the interface design of different entities.

## 4 Entities and Interfaces

### 4.1 SmartLobby

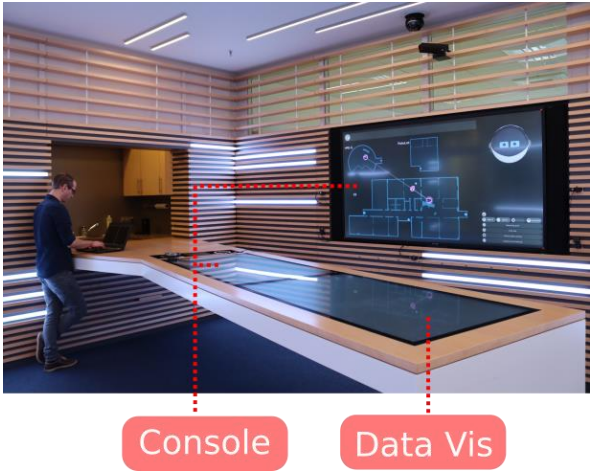

**Figure 3: SmartLobby entity. The console interface and data visualization are displayed on the three screens.**

SmartLobby is a lobby space equipped with various sensors, e.g. Kinect[2] cameras and microphone array, which can detect faces and locations of people in the room (Figure 3). Smartlobby has the capabilities of receiving command and informing people about the progress of the current task. The interface of SmartLobby consists of:

1. A GUI based interface. One large screen on the wall and two touch-screen tables were used to show the graphical interface (Figure 3). The left screen table and screen on the wall show the console interface of ICPS (as Figure 2 shows). User can also send commands through the touch-screen table. The right touch-screen table displays the data visualization of the Knowledge Graph (Figure 4).
2. A virtual robot face. A 3D virtual avatar robot. The virtual robot is shown on the right of the screen in the wall (Figure 5 right).
3. A speech interaction system. Users can give verbal commands, and then the system is able to recognize the intent of the person and provide feedback by speech.

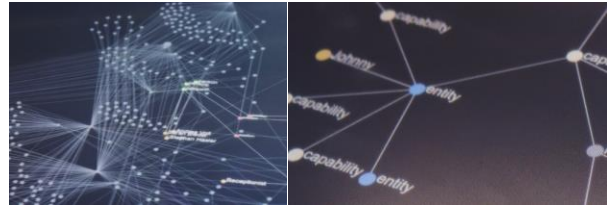

**Figure 4: The data visualization of the Knowledge Graph, which is displayed on the right touch screen table.**

### 4.2 Receptionist

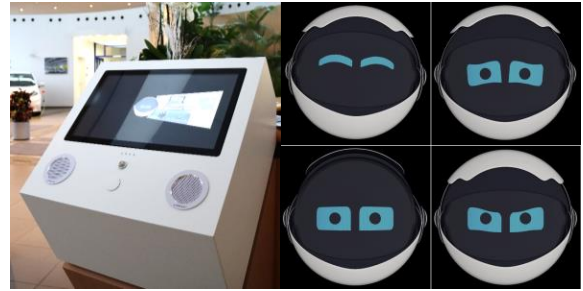

**Figure 5: Left: Receptionist entity used a virtual avatar robot to interact with humans during the registration phase. Interaction can be based on natural speech or buttons on screen depending on the situation. Right: Some facial expressions of the virtual robot.**

The ICPS includes a stationary computer showing a virtual receptionist (similar to the 3D virtual avatar as in the *SmartLobby*). The stationary computer is equipped with touch-screen, camera, microphones and it is used for registering new visitors to the system. After the registration process, visitor's information (name, face recognition model …) are stored in the *ICPS* backend knowledge representation. The interface of receptionist consists of:

1. A GUI based interface (Figure 5 left).
2. The 3D virtual avatar robot with a speech-interaction dialogue system. The virtual avatar and speech interaction is the same as the one in SmartLobby

---

[1] http://wiki.ros.org/multimaster_fkie

[2] https://developer.microsoft.com/en-us/windows/kinect

## 4.3 Mobile Robots

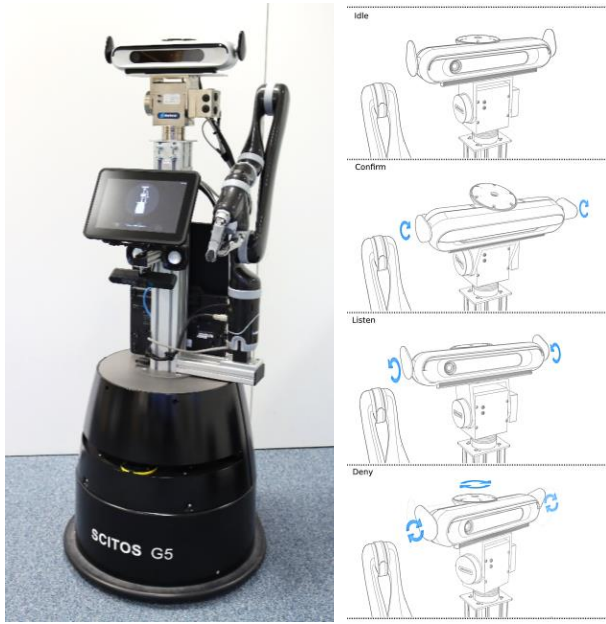

**Figure 6. Left: Mobile Robot Johnny. The interface of Johnny includes a physical robot head, a touch screen and a speech interaction system. Right: The robot head has ears and eyelid, which can express various gestures.**

We are using mobile robots based on MetraLabs' SCITOS G5[3]. They are additionally equipped with Kinect cameras mounted on a Schunk PW70[4] pan-tilt-unit for moving the head (Figure 6 left). The cameras are encased in a 3D-printed robot head. Laser scanners in the front and rear allow the robots to localize themselves in the room and Kinova JACO2 arms[5] enable the robots to transport objects. These robots have the capabilities of "moving to a specific location", "fetching an object" and "informing person" etc. They are equipped with sensors measuring their own pose and recognizing persons in 3 meters. The interface of mobile robots includes:

1. Speech interaction system. Similar to the virtual avatar in SmartLobby, users can interact with the robots with natural language. Each robot has its own voice. For example, Johnny and Walker have male voices and Ira has a female voice.
2. GUI. A 10-inch touch screen is attached on each robot's body, which is used for showing status information, assisting speech interaction and recognition of objects etc.
3. Physical robot head. A 3D-printed robot head is implemented in each robot, which includes eyelid and ears. A motion library of the head enables the physical head to deliver non-verbal signals by various pre-set gestures. Such as confirmation, denying or listening (Figure 6 right).

## 5 Example of use: Fetching Objects

This section we elaborate on one of the basic use cases implemented with the ICPS system: fetching_object. To be noted, there are more use cases of the ICPS system. For example, searching_for_person, which enables a user to know a person's location if any entity sees him/her; welcoming, which enables a guest to register in the reception, and lead him/her to a person or a room. we now detail the use case of fetching_object.

### 5.1 Receiving the command and planning

Markus (an alias) is in the smart lobby, he wants to have the smart safety belt (Figure 7). So he speaks to the avatar on the screen by natural languages, such as "I want the smart safety belt". SmartLobby entity receives the command from Markuss. Then the speech is converted to text, sent to the backend and analyzed by an intent recognition system. The backend sets a goal "Person_has smart safety belt". Currently, Markus is in the lobby, and his face is recognized by SmartLobby entity. The smart safety belt is in the secretary office. A robot-entity saw Andrea in the office 30mins ago. After aggregating the above information, the backend makes a plan in less than 2 seconds. The plan includes nine steps as follows:

1. Request control of entities.
2. Backend invokes Johnny
3. Johnny moves to the secretary office.
4. Johnny invokes Andrea.
5. Andrea fetches the key.
6. Andrea gives the key to Johnny.
7. Johnny moves to the Smart Lobby.
8. Johnny gives the key to Markus.
9. Release control of entities.

---

[3] https://www.metralabs.com/en/mobile-robot-scitos-g5/
[4] https://schunk.com/de_en/gripping-systems/series/pw-v6/
[5] https://www.kinovarobotics.com/en/products/robotic-arm-series

Then the avatar tells Markus by speech: "Plan generated. Johnny is sent to the secretary office to fetch the smart safety belt." Besides, taskbar on the screen shows the goal of the plan and progress of the executing. User can also stop executing by pressing the cancel button on the screen table. The entities, which are involved in the plan, are highlighted on the map.

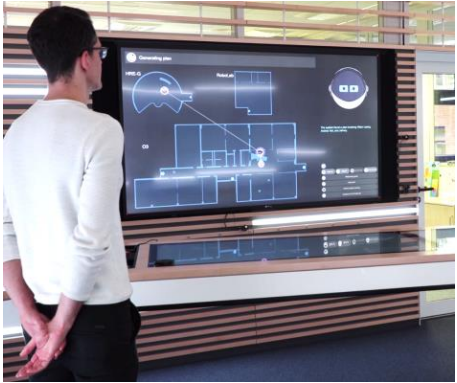

**Figure 7: Markus sends request in the lobby.**

## 5.2 Executing: searching for Andrea in the secretary office

When Johnny is invoked by the backend to execute the task, it shows notification gesture: left and right ears move back and forth twice alternatively. Furthermore, the text "executing the task of backend: go to the secretary office" is displayed on the screen of SmartLobby. Then Johnny starts to move to the secretary office. After it arrives at the room, it starts to turn its head around to search for Andrea while saying "Andrea?" Andrea should be recognized if she is nearby as her facial features are stored in the database. Andrea can also confirm that she is here by clicking the button on GUI.

## 5.3 Re-planning: adapting to the varying situation

In this use case, Andrea is not in the office now, even though she was seen by one entity 30mins ago. As a result, the original plan cannot be executed. Then the backend re-plans to let Johnny find an alternative person who is in the secretary office. Johnny sees that Sarah is also in the office now and she can open the storage to fetch the belt. Then Backend changes the plan to *asking Sarah for it*. The whole re-planning process should also be known by the user. The avatar in the *SmartLobby* tells Markus by speech: "cannot find the person in charge of the key, the plan changed. Searching for another person. Another person is found, a new plan is set." At the same time, Johnny asks Sarah: "*Hello Sarah, could you give me the smart safety belt?*" Sarah confirms by speech or clicks the button on the robot. Then robot Johnny stretches its arm and says "*Please put the smart safety belt in my hand, and press the confirm button on my screen.*" (Figure 8) After the handover of the belt, *Johnny* says "*Thank you!*" with a facial expression of appreciation: nodding, ears moving forwards and back and blinking eyelid once.

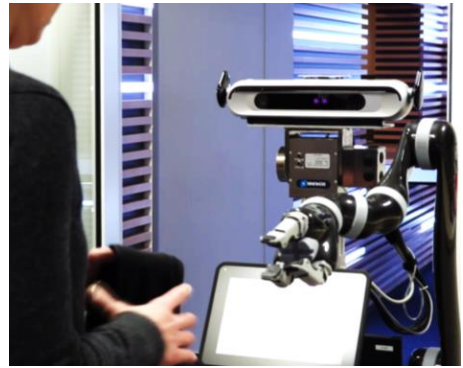

**Figure 8: Johnny asks another secretary for the smart safety belt.**

## 5.4 Giving the key to Markus

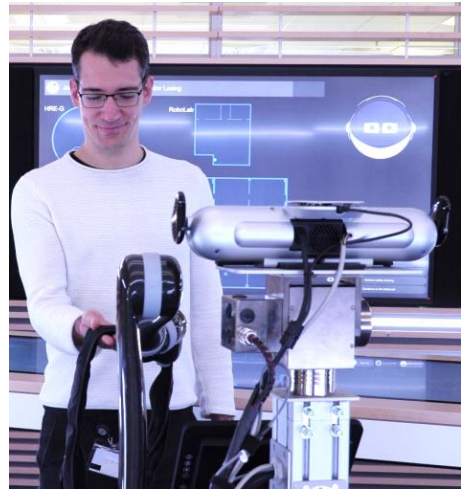

**Figure 9: Johnny gives the belt to Marcus**

After getting the key, Johnny goes back to the lobby. As the camera in the lobby recognizes Markus' face, Johnny knows the position of Markus and moves to the front of him. Then it says to Markus "Hello Markus, here is the key". Markus takes the key and presses the confirm button on Johnny's screen (Figure 9). Finally, the virtual robot of *SmartLobby* finishes the activity by saying "Goal is achieved."

## 6 Discussion and Future Work

In this paper, we presented a system architecture and the corresponding human-machine interaction design for a multi-entity intelligent system in the office environment. The interaction design pattern is different from the traditional single-entity robot system. There are four main advantages of our interaction design. First, our design provides an explainable and transparent interface to communicate the planning and executing process to users, and afford user involvement in different stages of this process. For example, users can keep track of the current goal, or progress of task execution, the location, status and next waypoint of different entities through a single screen in the *SmartLobby*. Second, the interaction pattern of each entity is holistic as well as characteristic. For example, the GUI style of *SmartLobby* and *Johnny* are highly coherent and consistent. At the same time, the avatar has its own voice (in a female tone) and richer facial expression; and *Johnny* has its own voice too (in a male tone) and a dedicated library of bodily posture and movement. People can clearly understand that different entities have distinguished responsibilities and capabilities. Third, both verbal communication channel (e.g. GUI interface and speech) and non-verbal communication channels (such as various gestures, facial expressions or audio feedback) are implemented in the system. Various interaction modalities complement each other to enhance the efficiency and user experience.

As for all such systems as presented in this paper, there are loose ends that allow for iterations in future work. This includes improving the planning method towards allowing for parallel execution of entity capabilities as well as more flexibly coping with uncertainties. Furthermore, the usage of the semantic knowledge representation would allow for incorporating more external world knowledge. And finally, the system is to be evaluated during long-term operation in this office environment.